\documentclass[twocolumn,aps,prd,superscriptaddress,nofootinbib]{revtex4}

\usepackage{graphicx}
\usepackage{amsmath,amsfonts}
\usepackage[hypertex]{hyperref}

\newcommand{\nn}{\nonumber}
\newcommand{\beq}{\begin{equation}}
\newcommand{\eeq}{\end{equation}}
\newcommand{\bea}{\begin{eqnarray}}
\newcommand{\eea}{\end{eqnarray}}

\newcommand{\GeV}{\mathrm{GeV}}

\newcommand{\ov}{\overline}
\newcommand{\one}{\ensuremath{\mathbf{1}}}
\newcommand{\three}{\ensuremath{{\mathbf{3}}}}
\newcommand{\threebar}{\ensuremath{\mathbf{\ov 3}}}
\newcommand{\six}{\ensuremath{\mathbf{6}}}
\newcommand{\eight}{\ensuremath{\mathbf{8}}}

\newcommand{\qqbar}{\ensuremath{{q\bar q}}}
\newcommand{\uubar}{\ensuremath{{u\bar u}}}
\newcommand{\ccbar}{\ensuremath{{c\bar c}}}
\newcommand{\ttbar}{\ensuremath{{t\bar t}}}
\newcommand{\mttbar}{\ensuremath{m_{t\bar t}}}
\def\d{{\rm d}}

\makeatletter
\def\simgt{\mathrel{\lower2.5pt\vbox{\lineskip=0pt\baselineskip=0pt
           \hbox{$>$}\hbox{$\sim$}}}}
\def\simlt{\mathrel{\lower2.5pt\vbox{\lineskip=0pt\baselineskip=0pt
           \hbox{$<$}\hbox{$\sim$}}}}
\makeatother

\begin{document}

\title{\boldmath Explaining the \ttbar\ forward-backward asymmetry without
dijet or flavor anomalies}

\author{Zoltan Ligeti}
\affiliation{Ernest Orlando Lawrence Berkeley National Laboratory,
University of California, Berkeley, CA 94720}

\author{Gustavo Marques Tavares}
\affiliation{Physics Department, Boston University, Boston, MA 02215}

\author{Martin Schmaltz}
\affiliation{Physics Department, Boston University, Boston, MA 02215}

\begin{abstract}

We consider new physics explanations of the anomaly in
the \ttbar\ forward-backward asymmetry measured at the Tevatron,
in the context of flavor conserving models.  The recently measured
LHC dijet distributions strongly constrain many otherwise
viable models.  A new scalar particle in the \threebar\
representation of flavor and color can fit the \ttbar\ asymmetry
and cross section data at the Tevatron and avoid both low- and
high-energy bounds from flavor physics and the LHC.
An $s$-channel resonance in $uc \to uc$ scattering 
at the LHC is predicted to be not far from the current sensitivity.
This model also predicts rich top quark physics for the early LHC from
decays of the new scalar particles. Single production gives
$\ttbar j$ signatures with high $p_T^{\rm jet}$, pair production
leads to $\ttbar jj$ and 4 jet final states.

\end{abstract}

\maketitle

\section{Introduction}

The unexpectedly large forward-backward asymmetry in the production of \ttbar\
pairs at the Tevatron as observed by CDF $A^\ttbar = 0.193 \pm
0.069$~\cite{Aaltonen:2008hc} and D\O\ $A^{\ttbar\ } = 0.24 \pm
0.14$~\cite{Abazov:2007qb} in 2008 generated a lot of interest, because it is
significantly higher than the Standard Model (SM) prediction, $A^{\ttbar\, \rm
(SM)} \approx 0.06$~\cite{Antunano:2007da, Bowen:2005ap, Kuhn:1998kw,
Almeida:2008ug}. One reason for being excited about this measurement is that the
top quark is a very sensitive probe of putative new physics at the TeV scale,
because of its large mass and coupling to the Higgs. Therefore, one might expect
signs of new physics to first show up in top physics. This hope has received a
boost by the recent CDF analysis, which showed that the asymmetry arises from
\ttbar\ events with high invariant masses~\cite{Aaltonen:2011kc}
\bea\label{CDFasym}
A^\ttbar(m_\ttbar > 450\,{\rm GeV}) &=& 0.475 \pm 0.114 \,, \nn \\ 
A^\ttbar(m_\ttbar < 450\,{\rm GeV}) &=& -0.116 \pm 0.153 \,.
\eea
The updated D\O\ result, only available integrated over \mttbar, and
uncorrected for effects from reconstruction or selection, $A^\ttbar = 0.08 \pm
0.04$~\cite{D0afb}, is consistent with the integrated CDF result, $A^\ttbar =
0.158 \pm 0.075$~\cite{Aaltonen:2011kc}.  So is the recent $A^\ttbar = 0.417 \pm
0.157$ measurement~\cite{CDFdilepton} in the dilepton channel, in which the raw
asymmetries, binned as in Eq.~(\ref{CDFasym}), also support the same trend. The
physics responsible for this anomaly may be related to CDF's high $p_T$ excess
in a boosted top search~\cite{CDFboostedtop}.  The large asymmetry at high
masses points towards tree-level exchange of a new heavy particle with strong
couplings to first and third generation quarks~\cite{Jung:2009jz,
Frampton:2009rk, Shu:2009xf, Arhrib:2009hu, Dorsner:2009mq, Burdman:2010gr,
Cheung:2011qa, Bai:2011ed, Shelton:2011hq, Blum:2011up, Grinstein:2011yv}. For
fits of four-fermion operators to the asymmetry data
see~\cite{Degrande:2010kt,Jung:2009pi}.

In absence of flavor symmetries, new states at the TeV scale with strong
couplings to quarks are severely constrained by the agreement of a vast amount
of flavor physics data with the SM (meson-anti-meson mixing, $CP$ violation,
rare decays). We are therefore motivated to look for an explanation of the
\ttbar\ asymmetry from new states whose couplings (and masses) preserve the full
flavor symmetries of the Standard Model quarks along the lines of
Refs.~\cite{Bauer:2009cc, Arnold:2009ay, Grinstein:2011yv}.

To do so, we classify the new particles not only by their spin and gauge
charges, but also by their quantum numbers under the flavor symmetries $SU(3)_Q
\times SU(3)_U \times SU(3)_D$. Here $SU(3)_Q$ is the set of transformations
which rotate the three generations of left-handed quark doublets, $Q$, and
$SU(3)_{U/D}$ transformations rotate the right-handed quark singlets, $U/D$. For
simplicity, and because this leads to the nicest model, we focus on the case
where the new states couple only to right-handed up-type
quarks.\footnote{Coupling to up-type quarks is preferred because it accesses the
large up-quark parton distribution function, and even $SU(3)_Q$ symmetric
couplings to the quark doublets give rise to new flavor violation proportional
to CKM matrix elements.} Depending on whether the coupling is to two quarks or
to a quark and an anti-quark, the new states have quantum numbers of a
``diquark" with baryon number 2/3 or a ``noquark'' with baryon number 0. Under
$SU(3)_U$ flavor the new particles must transform in one of the irreducible
representations contained in
\beq\label{options}
\mbox{diquark:}\ \ \three \otimes \three =  \threebar \oplus \six\,, \quad
\mbox{noquark:}\ \ \three \otimes \threebar = \one \oplus \eight\,. \
\eeq

\begin{figure}[tb]
\includegraphics[width=.42\columnwidth]{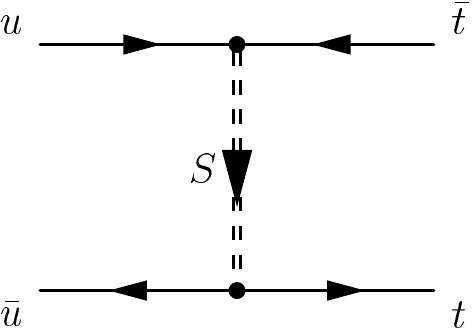} \hfil\hfil
\includegraphics[width=.42\columnwidth]{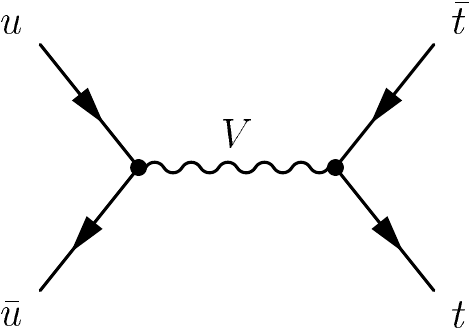}
\caption{Feynman diagrams for the new contribution to $u\bar u\to \ttbar$ from a
diquark (left) and a noquark (right).  The flavor symmetry implies an
additional $t$-channel diagram for the noquark~\eight\ (not pictured).}
\label{fig:diag}
\end{figure}

With regards to generating an asymmetry, the diquark models are nice because
diquarks contribute to \ttbar\ production in the $u$-channel (see
Fig.~\ref{fig:diag}). This new source of top quarks is peaked in the forward
direction and can easily produce a large asymmetry. The ``noquarks'' in the
flavor singlet representation are closely related to the extensively discussed
axigluons and do not provide a very good fit to the asymmetry data. The main
problem is that they are $s$-channel resonances coupling to up quarks and to top
quarks. They would give rise to features in the invariant mass distribution $d
\sigma_\ttbar/d m_\ttbar$ of \ttbar\ pairs and also of dijets at the Tevatron
and the LHC~\cite{Bai:2011ed}. The case of the \eight\ of flavor is more
interesting as it contributes to \ttbar\ production in the $t$-channel and the
$s$-channel. It is possible to find good fits to both the asymmetry as well as
the total \ttbar\ cross section in this case for either light ($M_8\sim
300$~GeV) or heavy ($M_8 \sim 1000$~GeV) new states~\cite{Grinstein:2011yv}.
Data on dijet resonances from the Tevatron~\cite{Aaltonen:2008dn, CDFchidist,
Abazov:2009mh} and SPS~\cite{Arnison:1986np, Alitti:1990aa} rule out flavor
universal couplings in the light case. In addition, light spin one particles are
associated with gauge symmetries. This would imply at least an approximate gauge
symmetry of flavor, which would need to be broken by the mechanism generating
the up-type Yukawa couplings. The heavy state requires very large couplings to
generate a large enough asymmetry and can be shown to violate recent bounds on
dijets from the LHC~\cite{Khachatryan:2011as}.\,\footnote{The dijet bounds can
be evaded with large flavor breaking in the third generation
couplings~\cite{Grinstein:2011yv}.} 
We will not consider the \one\ or \eight\ any further.

Concentrating on renormalizable interactions, the \threebar\ and \six\ must be
complex scalar fields and their couplings to quarks can be written as
\beq\label{coupling}
\lambda\, U^{a\alpha}\, U^{b\beta} (S^*)^{r\rho}\, T_{rab}\, T_{\rho\alpha\beta}\,.
\eeq
Flavor indices $a,b$ on the quarks run from 1 to 3 and the flavor index on $S$
runs over $r=1\ldots3$ or $r=1\ldots6$. We also explicitly displayed the color
indices $\alpha,\beta=1\ldots3$ on the quarks
and $\rho=1\ldots3$ or $\rho=1\ldots6$ on $S$.
Note that the color and flavor quantum numbers of $S$ are correlated,
because the two identical quark fields must be symmetric under interchange of
color and flavor indices. The scalar \threebar\ is a $(\overline 3,1)_{4/3}$
under the Standard Model gauge group $(SU(3)_{c}, SU(2)_{w})_{U(1)}$ whereas the
\six\ is a $(6,1)_{4/3}$. The invariant tensors $T$ are therefore the same for
color and flavor. $T_{rab}$ for the \threebar\ is antisymmetric in $a$ and $b$
whereas for the \six\ it is symmetric. We emphasize that in either case these
couplings preserve the $SU(3)_U$ flavor symmetry and do not contribute to flavor
violating processes.

An obvious but important consequence of the flavor symmetry is that it relates
processes involving quarks in different generations.
In particular, we observe an important distinction between the \threebar\ and
the \six. Equation~(\ref{coupling}) for the \six\ contains a coupling of two up
quarks to $S$. This leads to an $s$-channel resonance in $uu\to uu$
scattering, which accesses the largest high energy parton luminosities at the
LHC and is already severely constrained by recent dijet analyses from
CMS~\cite{Khachatryan:2010jd, Khachatryan:2011as} and
ATLAS~\cite{Collaboration:2010eza}. As we will show below, a measurement of
dijet angular distributions at CMS~\cite{Khachatryan:2011as} already rules out
the entire range of couplings and masses for the \six\ which generates an
appreciable \ttbar\ asymmetry. The \threebar\ is antisymmetric in its couplings
to quarks so that the coupling to two up quarks vanishes.

This leaves the scalar triplet as our only flavor preserving candidate for
explaining the \ttbar\ forward-backward asymmetry at the Tevatron. In the
following Sections we compare the predictions of this model to all relevant data

\begin{enumerate}\vspace*{-4pt}\itemsep 0pt

\item The binned \ttbar\ asymmetry in
Eq.~(\ref{CDFasym})~\cite{Aaltonen:2011kc};

\item The $p\bar p\to \ttbar$ total cross section, $\sigma_\ttbar = (7.5 \pm
0.48)$\,pb~\cite{CDFsigma};

\item Measurement of $\d\sigma_\ttbar/\d\mttbar$~\cite{Aaltonen:2009iz};

\item The recent CMS dijet analysis~\cite{Khachatryan:2011as}.

\end{enumerate}\vspace*{-4pt}

We find that unlike all the other models the triplet scalar is currently
unconstrained by dijet data. However, there is some tension between the large
\ttbar\ asymmetry required at high invariant masses and the shape of the
measured differential cross section as a function of the invariant mass of the
\ttbar\ pair. In particular, a 40\% asymmetry requires a new physics
contribution which is very asymmetric and comparable in size to the \ttbar\
cross section from QCD, but which does not significantly change the shape of the
cross section. We are not aware of any model in the literature which completely
accomplishes this.

The best fit to the \ttbar\ asymmetry and cross section in our model is obtained
for triplet masses in the range 500--800~GeV with relatively large couplings,
$\lambda \sim 1.5-3.0$ (for even larger couplings our perturbative calculations
quickly become suspect). It is therefore easily within reach of the LHC, and
both single and pair production of these states should be possible with large
cross sections. The most promising processes are single production with top
quarks $u g \rightarrow \bar{t} S_{tu} \rightarrow \ttbar j$ and pair production
$gg,\, \uubar \rightarrow S S^* \rightarrow jjjj,\, \ttbar jj$. In the context
of flavor-symmetric models it would be particularly interesting if one could
measure the forward-backward asymmetries in dijet events with pairs of high
$p_T$ charm or bottom quark jets at the
Tevatron~\cite{Bai:2011ed,Strassler:2011vr}. Our model predicts an asymmetry for
charm quarks similar to that for top quarks and vanishing new physics
contribution to the bottom quark asymmetry.

The remainder of the paper is organized as follows.  In Section~\ref{sec:model}
we define the \threebar\ and the \six\ models, compute the \ttbar\ cross section
and asymmetry in each and determine the preferred region in parameter space by
fitting to the Tevatron \ttbar\ data. We then compute the predicted dijet rates
at the Tevatron and LHC and show that the preferred region of the \threebar\
model is still allowed whereas the \six\ model is ruled out by the CMS dijet
measurement. In Section~\ref{sec:lhc} observable predictions for the LHC from
the diquark of the \threebar\ model are explored.

\section{The \threebar\ and the \six}
\label{sec:model}

The two models contain a new scalar in the \threebar\ or \six\ representation of
flavor $SU(3)_U$.  The interaction Lagrangian is
\beq\label{3def}
{\cal L} = \lambda\, U^{a\alpha}\, U^{b\beta} (S^*)^{r\rho}\,
  T_{rab}\, T_{\rho\alpha \beta} \, + {\rm h.c.}\,,
\eeq
where Latin (Greek) indices denote flavor (color). The fields $U$ are the
right-handed up-type quark singlets with SM charges $(3,1)_{2/3}$, their Lorentz
indices are contracted with $i\sigma_2$. The invariant tensor for the \threebar\
is $T_{rab} = \epsilon_{rab}/\sqrt{2}$. For the \six\ the tensor $T_{rab}$ can
be decomposed into six $3\times 3$ real symmetric matrices, $T_r$. In computing
amplitudes from exchange of these particles we will use the identities
\bea
(T_3)_{rab}\, (T_{3})^{rcd} &=&
  \frac12 \big(\delta^c_a \delta^d_b - \delta^c_b \delta^d_a\big)\,, \nn\\
(T_6)_{rab}\, (T_6)^{rcd} &=&
  \frac12 \big(\delta^c_a \delta^d_b + \delta^c_b \delta^d_a\big)\,,
\eea
which also fix the normalization of our invariant tensors. Here $T^{rcd} \equiv
T_{rcd}^* = T_{rcd}$, because we defined the tensors to be real. Each model has
only two parameters, the coupling $\lambda$, and the mass $m_S$. Our definition
of the coupling $\lambda$ differs by a factor of $\sqrt{2}$ in normalization
compared to the coupling $y$ of Ref.~\cite{Shu:2009xf}, so that
$\lambda=y/\sqrt{2}$.
Of course, other viable models may be obtained by including additional free
parameters. In the context of minimal flavor violation one might be motivated to
consider insertions of $Y_UY_U^\dagger$. These  would have very small effects on
light quarks but could change the top quark couplings significantly. We have
refrained from doing so in the  interest of simplicity and because we expect
insertions of $Y_UY_U^\dagger$ generated from loops to be small.

The process $u(p_1)\, \bar u(p_2) \to t(k_1)\, \bar t(k_2)$ is given in the SM
by $s$-channel gluon exchange. The $S$ interaction mediates a $u$-channel
contribution (Fig.~\ref{fig:diag}). Including both contributions, the
differential partonic cross section is~\cite{Shu:2009xf, Arhrib:2009hu,
Dorsner:2009mq}
\bea\label{dcostheta}
\frac{\d\sigma_\ttbar}{\d\cos\theta} &=&
  \frac14\, \frac19\, \frac\beta{2\pi s}
  \bigg[g_s^4\, \frac{t_t^2 + u_t^2 + 2 s\, m_t^2}{s^2} \nn\\
&&{} +  g_s^2\, \lambda^2\, C_0\, \frac{u_t^2+s\, m_t^2}{s\, u_S} 
  +  \lambda^4\, C_2\, \frac{u_t^2} {u_S^2} \bigg] .
\eea
Here $\theta$ is the scattering angle between the outgoing top and the incoming
quark in the partonic center-of-mass frame, and $\beta \equiv \sqrt{1 - 4m^2_t /
s}$. The Mandelstam variables are $s \equiv (p_1 + p_2)^2$, $t \equiv (p_1 -
k_1)^2$, $u \equiv (p_1 - k_2)^2$, and we denoted $t_X = t - m_X^2$ and $u_X = u
- m_X^2$ ($X=t,S$). Finally, the color factors are $C_0=1$ and $C_2=3/4$ in the
case of the \threebar, and $C_0=-1$ and $C_2=3/2$ for the \six.\,\footnote{Here
we note a typographical error in the sign of $C_0$ for the case of the \six\ in
Ref.~\cite{Shu:2009xf}, which was corrected in Ref.~\cite{Arhrib:2009hu}.}

\subsection{\boldmath Fitting the \ttbar\ asymmetry at the Tevatron}

Our strategy is to fit the models to the \ttbar\ related data from the Tevatron,
and then explore whether the resulting parameter space is consistent with other
experiments.  We do not perform a $\chi^2$ fit to the data, because the required
correlations are not available. One can get a reasonable understanding of the
``goodness of fit'' of the models by plotting the constraints from various
experiments as functions of the models' two parameters.

\begin{figure}[tb]
\includegraphics[width=1.0\columnwidth]{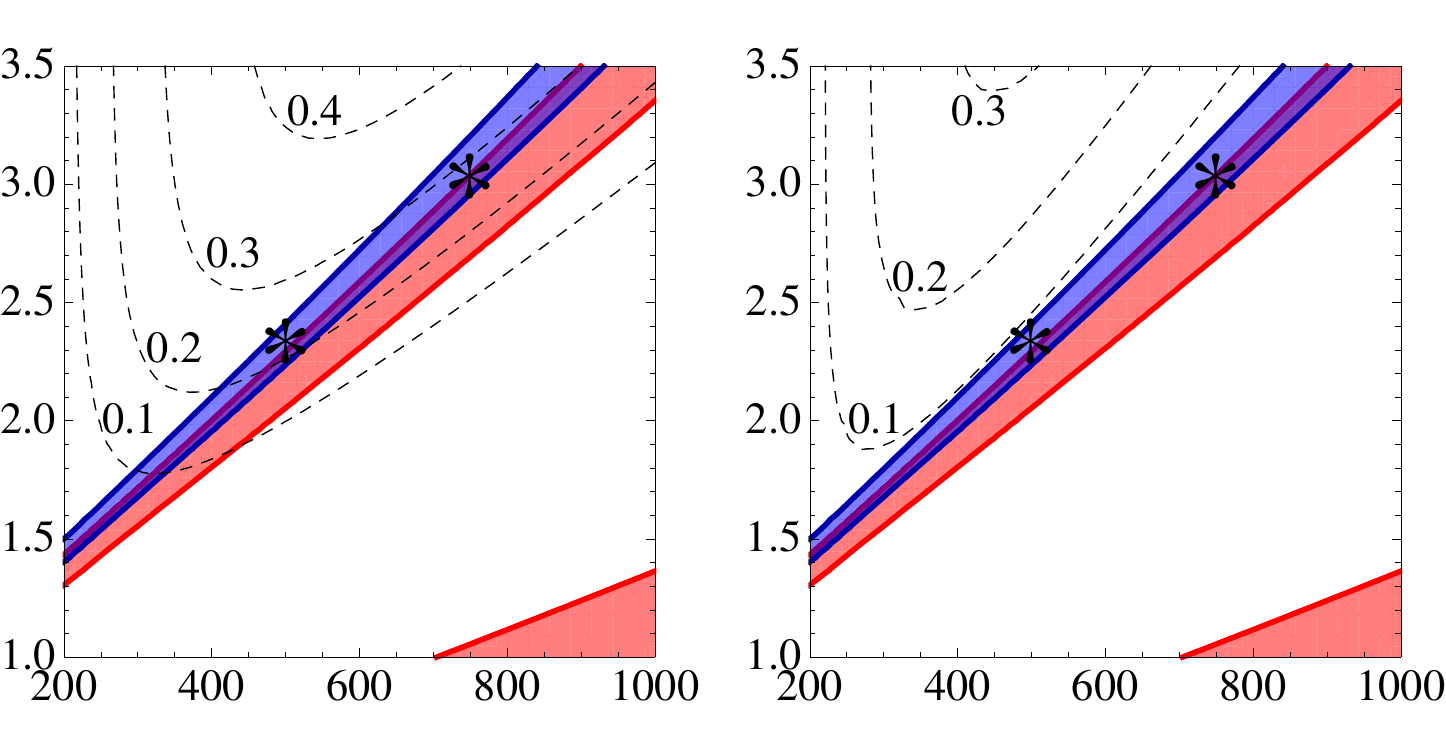} 
\caption{Regions in $m_S$ vs.\ $\lambda$ parameter space for the \threebar\
model which are allowed by the total cross section contraint. The red (lighter)
and blue (darker) shaded regions correspond to $\sigma_\ttbar -
\sigma_\ttbar^{\rm (SM)} = (0.0 \pm 0.7)$\,pb and $(1.0 \pm 0.7)$\,pb,
respectively.  The contours in the left [right] plot show the new physics
contribution to the \ttbar\ forward-backward asymmetry $A^{\ttbar}_{\rm
NP}(\mttbar > 450$~GeV) [$A^{\ttbar}_{\rm NP}(\mttbar < 450$~GeV)].  The total
asymmetry in each bin is obtained by adding the Standard Model contributions,
$A^{\ttbar}_{\rm SM}=0.09$ (left) and $0.04$ (right)~\cite{Aaltonen:2011kc}.}
\label{fig:threeplots}
\end{figure}

In agreement with previous work we find that the well-measured total cross
section, $\sigma(p\bar p \to \ttbar) = (7.5 \pm 0.48)$\,pb~\cite{CDFsigma}, 
provides the most important constraint on the parameter space. There is
currently some debate in the literature about the precise value of the
theoretical prediction. State of the art
NLO+NNL~\cite{Moch:2008qy,Cacciari:2008zb,Kidonakis:2008mu} calculations quote
about a 10\% uncertainty, and fit the data well. On the other hand, recent
calculations resumming threshold logs obtain lower values, around
6.5\,pb~\cite{Ahrens:2010zv}. Given this uncertainty in the predictions, we
choose a conservative approach. We plot the allowed regions corresponding to the
central values of each of the theory predictions, with approximately 10\% total
uncertainties, $\pm 0.7$\,pb. Thus, the NLO+NNL calculations allow the new
physics contribution to the \ttbar\ cross section to account for $\sigma_\ttbar
- \sigma_\ttbar^{\rm (SM)} = (0.0 \pm 0.7)$\,pb, and yield the red (lighter)
shaded allowed regions in Figure~\ref{fig:threeplots} for the \threebar\ model.%
\footnote{In our calculations of the QCD and new physics contributions we
applied a K-factor of $1.3$. We use the CTEQ-5L parton distribution functions~\cite{Lai:1999wy} implemented in Mathematica, and checked that MSTW 2008~\cite{Martin:2009iq} gives compatible results.}
Using the threshold resummed predictions, there is additional room for new
physics contributions to the \ttbar\ cross section, and the blue (darker) shaded
regions show $\sigma_\ttbar - \sigma_\ttbar^{\rm (SM)} = (1.0 \pm 0.7)$\,pb. In
the former case there are two allowed regions in parameter space. The less
interesting region is near the Standard Model and has small Yukawa couplings and
therefore small effects on the forward-backward asymmetry. The regions of
interest correspond to narrow bands in parameter space with larger Yukawa
couplings. This region is consistent with the total cross section constraint
because of a cancellation: the new physics squared contribution to the cross
section cancels against the interference with the QCD contribution. Since the
total cross section is the most precise of the measurements, the allowed
parameter space is largely defined by these narrow bands.

Overlaid on the same plot are contours of constant predicted \ttbar\
forward-backward asymmetry in the high invariant mass bin, $\mttbar > 450$~GeV.
One sees that our model can generate parton level \ttbar\ asymmetries between
20\% and 30\% from the new physics alone. The standard model contributes an
additional asymmetry of 0.09 in this bin~\cite{Aaltonen:2011kc}. To a reasonable
approximation the two contributions can simply be added, and  the combined
asymmetry overlaps the $1\,\sigma$ preferred region of the CDF measurement
$A^\ttbar(m_\ttbar > 450\,{\rm GeV}) = 0.475 \pm 0.10 \pm 0.05$. In the right
panel of Fig.~\ref{fig:threeplots} we show contours of constant forward-backward
asymmetry at low invariant masses, $\mttbar <450$~GeV, overlaid with the same
regions allowed by the total cross section constraint. One sees that even though
a large asymmetry of 20--30\% is generated at high invariant masses, the
asymmetry at low invariant masses is always less than 10\% in the region of the
parameter space allowed by the total cross section constraint. Based on these
``eyeball'' fits we define two benchmark points which provide reasonable fits to
the CDF asymmetry and total cross section,
\bea\label{studypoints}
&& \mbox{(1) ``low mass":}\quad\ m_S = 500\,\GeV,  \quad\  \lambda = 2.3\,,
  \qquad\nn\\
&& \mbox{(2) ``high mass":}\quad m_S = 750\,\GeV,  \quad\  \lambda = 3.0\,.
\eea

\begin{figure}[tb]
\includegraphics[width=1.0\columnwidth]{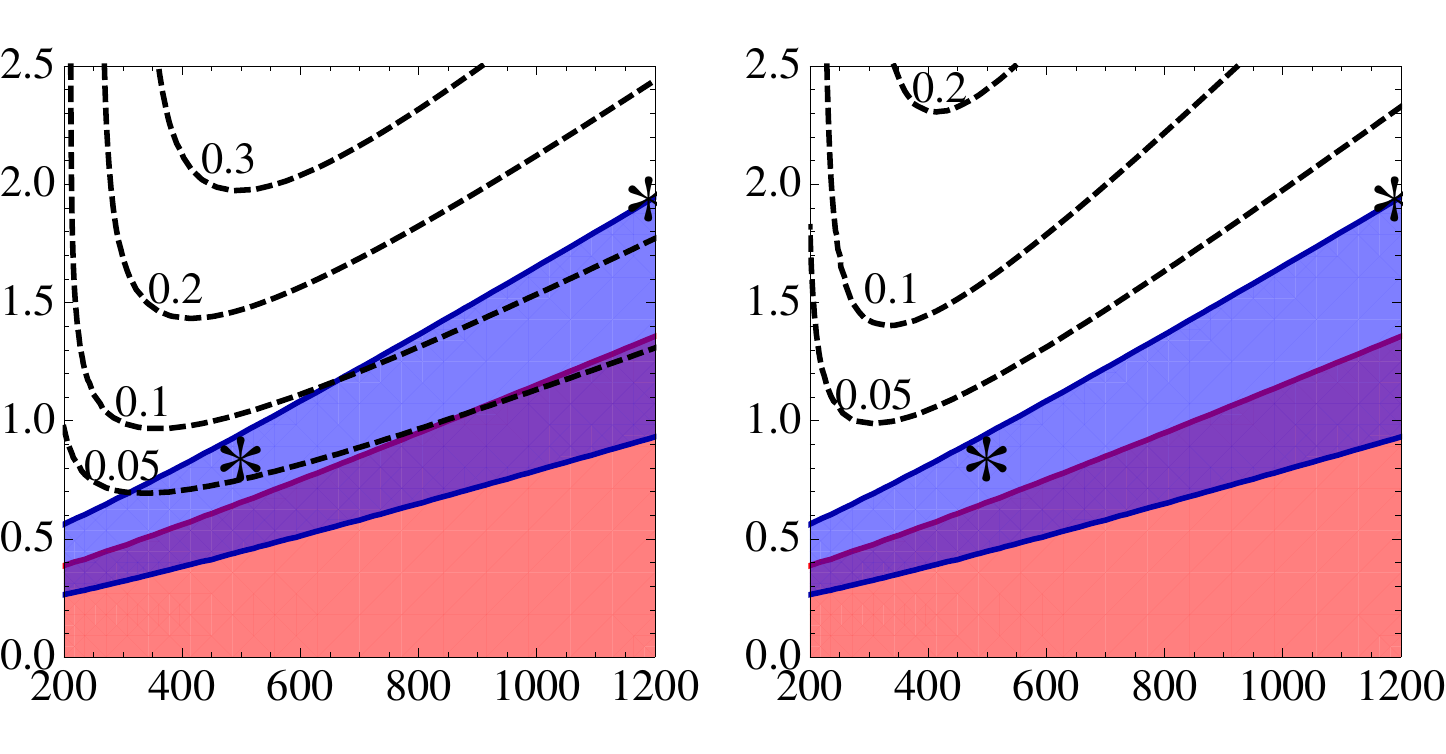} 
\caption{The contour plots as in Fig.~\ref{fig:threeplots}, but for the \six\
model.}
\label{fig:sixplots}
\end{figure}

The same fit result for the \six\ model is shown in Fig.~\ref{fig:sixplots}. 
This model cannot accommodate as high \ttbar\ forward-backward asymmetries as
the \threebar. Moreover, as discussed below, the dijet constraints already rule
it out. For definiteness we also define two benchmark points for this model with
$(m_S=1200\ {\rm GeV},\ \lambda=1.9)$ and $(m_S=500\ {\rm GeV},\ \lambda=0.8)$.
The two points generate high invariant mass ($\mttbar >$ 450 GeV) asymmetries
of 12\% and 6\%, respectively. 

CDF also provided a binning of the \ttbar\ asymmetry by the rapidity difference
between the top anti-top pair. There is a large asymmetry at large rapidity
difference and a small asymmetry at small rapidity difference. We find that in
our model this binning does not generate new constraints on the parameter space.
Our model is consistent with the rapidity binned data within the sizable
$1\,\sigma$ errors quoted in Ref.~\cite{Aaltonen:2011kc}.

In Table~\ref{tab:studypoints} we summarize the predictions of the two benchmark
points in comparison to the measurements. Our predictions for the asymmetry
include both new physics and standard model
contributions~\cite{Aaltonen:2011kc}.

\begin{table}[b]
\begin{tabular}{c|c|c c}
Observable  &  Measurement  &  ~Point\,(1)~  &  Point\,(2) \\
\hline\hline
$A^\ttbar(m_\ttbar > 450\,{\rm GeV})$~ & $0.475 \pm 0.114$ & 0.30  &  0.36 \\ 
$A^\ttbar(m_\ttbar < 450\,{\rm GeV})$ & ~$-0.116 \pm 0.153$~ & 0.10  &  0.07 \\
\hline
$A^\ttbar(|\Delta y| \geq 1)$  &  $0.611 \pm 0.256$  &  0.42  &  0.46 \\ 
$A^\ttbar(|\Delta y| < 1)$  &  $0.026 \pm 0.118$  & 0.12  &  0.12 \\
\hline
$\sigma_\ttbar - \sigma_\ttbar^{\rm (SM)}$  &  \mbox{see the text}
  &  0.7\,pb  &  0.5\,pb \\ 
\hline\hline
\end{tabular}
\caption{Comparison of the CDF binned asymmetry
measurements~\cite{Aaltonen:2011kc} with the benchmark points in
Eq.~(\ref{studypoints}) for the \threebar\ model.}
\label{tab:studypoints}
\end{table}

Finally, the shape of the \ttbar\ cross section, $\d\sigma_\ttbar / \d\mttbar$,
has also been measured~\cite{Aaltonen:2009iz}. The SM prediction for this
spectrum is also known at NLO+NNL~\cite{Moch:2008qy, Cacciari:2008zb,
Kidonakis:2008mu}. However, the theoretical uncertainties are larger than for
the total cross section, especially at large \mttbar~\cite{Ahrens:2010zv}.  We
do not perform a fit to the spectrum, and only compare the prediction of our
model for the cross section in a high invariant mass bin, 700 GeV$< \mttbar <$
800 GeV, following~\cite{Blum:2011up}. This bin is fairly far from the bulk of
the \ttbar\ data, and therefore tests a different region of the \ttbar\ spectrum
than the total cross section.  In addition, the cross section at the highest
invariant masses is expected to be the most sensitive to the new physics
contributions.  We find that there is significant tension between the measured
cross section and the model prediction for $\sigma_\ttbar - \sigma_\ttbar^{\rm
(SM)}$ in this bin, with the latter being about twice the SM prediction (a
similar excess is found in other models in the literature). Given that both
theoretical and experimental uncertainties are substantial for the  tail of the
\ttbar\ spectrum, we set this issue aside and explore what new information can
be obtained from LHC experiments in the context of this model.

\subsection{Dijet constraints}

We next study the dijet constraints on the \threebar\ model and contrast them
with the corresponding constraints for the \six. Since the coupling of the
\threebar\ in flavor space is $\epsilon_{rab} U^a U^b$ it does not mediate
$uu\to uu$ scattering. This is fortunate because a scalar $s$-channel resonance
of the leading $uu$ parton luminosity with coupling $\lambda > 1$ can be ruled
out for masses $\sim$~0.4--3~TeV with the recent CMS analysis of dijet angular
distributions~\cite{Khachatryan:2011as}. The \threebar\ model does predict an
$s$-channel dijet resonance in $uc \to uc$ scattering. As we will see the
sensitivity of the CMS analysis is close to what would be required to discover
it. The model also gives rise to dijets from  $u$-channel $\uubar\to \ccbar$
processes both at the LHC and at the Tevatron. However, this process is less
sensitive than $\uubar\to \ttbar$ or $uc \to uc$.

The CMS collaboration measured the differential dijet cross section $\d \sigma /
\d\chi$, where $\chi = (1+|\cos\theta|) /(1-|\cos\theta|)$ is defined such that
it makes the QCD prediction for $\d \sigma / \d\chi$ flat at small $\theta$. In
our models we predict
\bea\label{dsigmadchi}
\frac{\d\sigma}{\d\chi} &=&
  \frac{1}{36\pi s (1+\chi)^2} \bigg[ g_s^4 \left( \frac{s^2+u^2}{t^2}
  + \frac{s^2+t^2}{u^2} -\delta\, \frac{2s^2}{3ut} \right) \nn\\
&&{} + D_0\, g_s^2\, \lambda^2\left( \frac{s}{t} + \frac{s}{u} \right)
  \frac{s(s-m_S^2)}{(s-m_S^2)^2 + m_S^2\, \Gamma_S^2} \nn\\
&&{} + 3 D_2\, \lambda^4 \frac{s^2}{(s-m_S^2)^2 + m_S^2\, \Gamma_S^2} \bigg]\,.
\eea
For $u u \rightarrow u u$ in the \six\ model we have $\delta = 1$, $D_0=-2$, and
$D_2=2$, while for $uc \rightarrow uc$ in the \threebar\ model $\delta = 0$,
$D_0=1$, and $D_2=1/2$. In both models the width of the resonance is $\Gamma_S =
m_S\, \lambda^2/(8\pi)$.  Note that $\chi$ does not distinguish between $\pm
\cos\theta$, and consequently when computing $\d \sigma / \d\chi$ for processes
where the final state is composed of distinguishable particles, we summed the
contributions of both signs of $\cos\theta$. 

The CMS analysis contains 9 different dijet invariant mass regions, from
250--350~GeV to above 2.2~TeV. For optimal sensitivity to $s$-channel new
physics beyond the standard model, one looks for a rise in the number of events
in the most central bin, $1<\chi<2$. QCD predicts an approximately flat
distribution over all $\chi$ bins with a small rise in the central bin.  The
measured cross section in each bin is normalized to the total dijet cross
section sumed over all $\chi$ bins.  The remaining uncertainty in the central
$\chi$ bin is estimated by CMS to be about 10\%~\cite{Khachatryan:2011as}.

\begin{figure}[t]
\includegraphics[width=\columnwidth,clip]{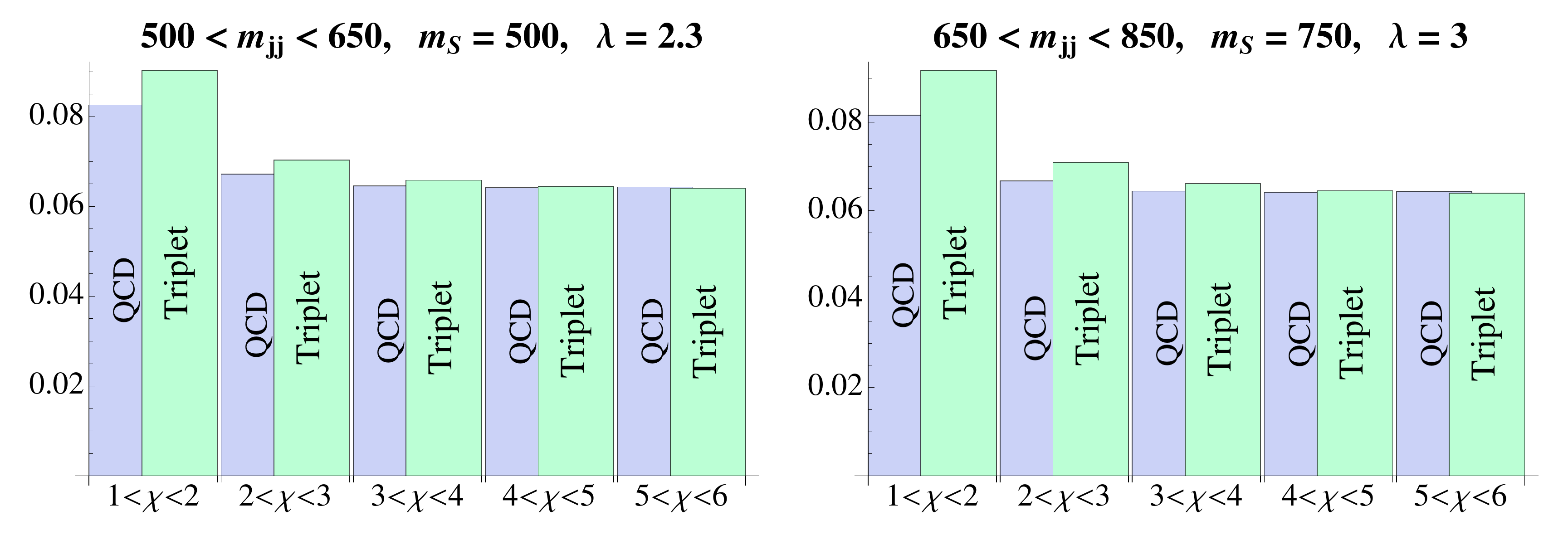} 
\caption{Normalized dijet cross sections in $\delta\chi=1$ bins for $1<\chi<6$. 
The blue (darker) histogram shows the QCD prediction, while the green (lighter)
histogram shows the normalized sum of QCD and the new  physics $uc\to uc$
contribution to the dijet rate in the \threebar\ model.  The left and right
plots show $500\,\GeV < m_{jj} < 650\,\GeV$ and $850\,\GeV < m_{jj} <
1100\,\GeV$, respectively. These are the mass ranges for which the largest
excess above QCD is predicted for the two benchmark points (1) and (2) in
Eq.~(\ref{studypoints}). In either case the excess in the most central bin is
about 10\% which is comparable to the uncertainties quoted
in~\cite{Khachatryan:2011as}.}
\label{fig:dijet3}
\end{figure}

The scalar resonance in $uc\to uc$ scattering in the \threebar\ model does
predict an increased dijet rate at small $\chi$, most strongly for dijet
invariant masses near $m_S$. In Fig.~\ref{fig:dijet3} we show the predicted
normalized dijet rate $(1 / \sigma_{\rm dijet})\,(\d\sigma_{\rm dijet} /
\d\chi)$ for our model compared with QCD. For each of the two benchmark points
we show the first five $\chi$ bins for dijet invariant masses near $m_S$.  There
is a discernible rise in the cross section above QCD, but the systematic
uncertainties shown in Ref.~\cite{Khachatryan:2011as} are of the same size or
larger than the new physics effect in Fig.~\ref{fig:dijet3}.  However, increased
statistics combined with data driven background subtraction (at higher and lower
$m_{\rm dijet}$) may be sensitive to this $uc$ resonance in the future. If the
data starts to show signs of a resonance, it would be very interesting to employ
(even limited) charm tagging to confirm the presence of charm quarks.

\begin{figure}[t]
\includegraphics[width=\columnwidth,clip]{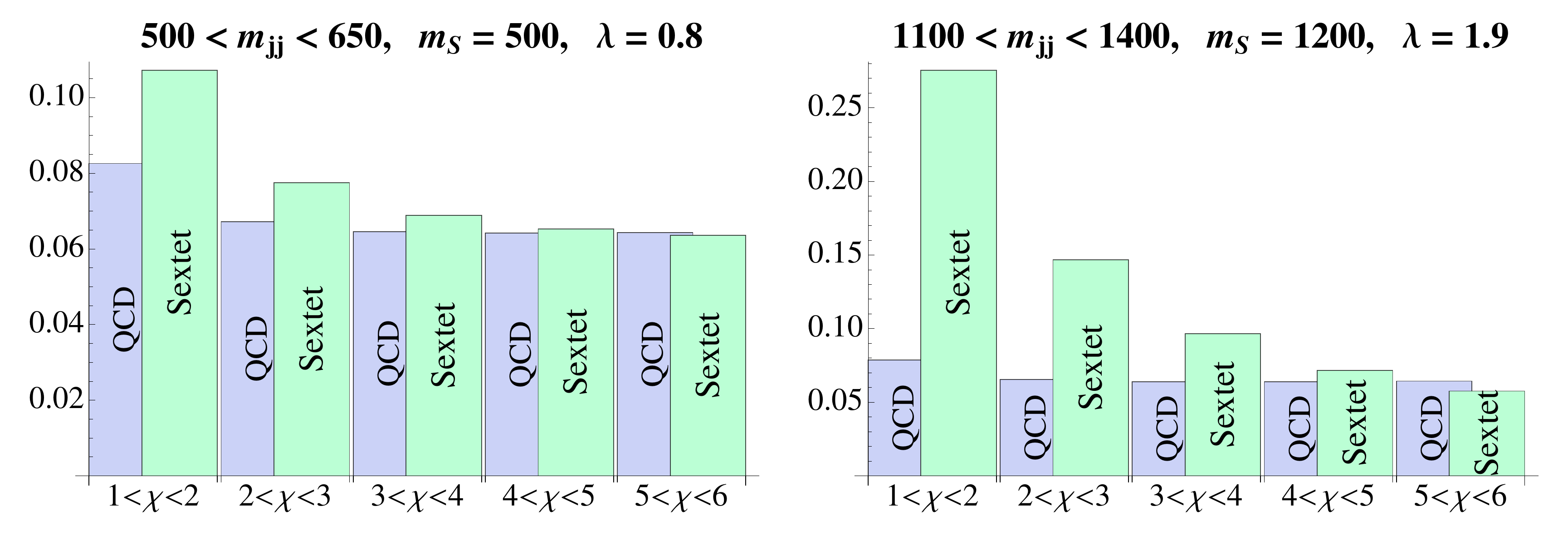} 
\caption{Same as Fig.~\ref{fig:dijet3}, but for $uu\to uu$ in
the \six\ model.  The left plot shows the contribution for $500\,\GeV < m_{jj}
< 650\,\GeV$, while the right plot shows that for $1100\,\GeV < m_{jj} <
1400\,\GeV$.}
\label{fig:dijet6}
\end{figure}

For comparison, we also show the predicted dijet $\chi$ spectrum for the \six\
model in Fig.~\ref{fig:dijet6}. Here the resonance is in $uu \to uu$ scattering,
and consequently the new physics signal is very large. We see that the heavy
benchmark point for the \six\ is ruled out by a large margin. The light
benchmark point predicts a rise of about 30\% beyond QCD which is also ruled
out.

\subsection{(Un)Naturalness: SUSY and the Landau pole}

Fundamental scalar particles such as our triplet suffer from a naturalness
problem, as their masses are quadratically sensitive to ultraviolet scales.
Therefore, one should think of this theory as the low energy limit of a more
complete theory in which the scalar arises as a composite. Alternatively, the
scalar mass may be made natural with supersymmetry. It is straightforward to
supersymmetrize our model and we briefly describe the resulting model here.

The minimal supersymmetric version of our model is the MSSM with one extra
chiral superfield $\overline S$ with identical gauge and flavor quantum numbers
as our scalar, and another chiral superfield $S$ with the opposite gauge and
flavor quantum numbers. The Yukawa coupling of Eq.~(\ref{3def}) is lifted to a
superpotential term $W = \lambda\, UU\!S$ where now $U$ and $S$ denote the full
superfields. Flavor and gauge indices are contracted as in the
non-supersymmetric theory. The mass of the scalar gets contributions from
supersymmetry breaking and supersymmetry preserving terms. If it primarily
arises from supersymmetry breaking one expects the $S+\overline S$ fermions to
be light. These fermions are R-parity odd, they can be produced in pairs at the
LHC or appear in cascade decays of squarks, giving events with high jet
multiplicities and missing energy. The resulting signatures are similar to the
ones of a light gluino.

A separate issue the reader may worry about is that the relatively large Yukawa
couplings invalidate perturbation theory and lead to a Landau pole not far from
the mass of the triplet. The leading term in the beta function for~$\lambda$~is
\beq
16 \pi^2\, \frac{\d \lambda}{\d(\ln \mu)} = 4\, \lambda^3 \,.
\eeq
We see that there are no large multiplicity factors associated with color and
flavor and the loop expansion parameter is approximately $\lambda^2/(4\pi^2)$,
which is perturbative for the couplings of interest. The solution to the
renormalization group equation is $1/\lambda^2(\mu) = \ln(\Lambda/\mu)/(2
\pi^2)$, where $\Lambda$ is the Landau pole.
For example, $\lambda(m_S)=2.3$ and
$m_S=500$~GeV gives $\Lambda \sim 21$~TeV.  (The other benchmark point in
Eq.~(\ref{studypoints}), $\lambda(m_S)=3$ and $m_S=750$~GeV, gives $\Lambda \sim
7$~TeV.) In either case the Landau pole is far enough that dimension-6 operators
suppressed by $16\pi^2/\Lambda^2$ are much smaller than the $S$ exchange
diagrams.

\section{\boldmath LHC signals: $uc$ resonance and $pp\to \ttbar j$}
\label{sec:lhc}

While \ttbar\ production at the Tevatron is dominated by $\qqbar \to \ttbar$,
the total $pp\to\ttbar$ cross section at the LHC is dominated by $gg\to\ttbar$,
which does not exhibit a forward-backward asymmetry.  By measuring \ttbar\ pairs
at higher average rapidity or higher invariant mass, one can enhance the \qqbar\
initial state, and thus possibly check the Tevatron observation.  Another
possibility is to measure the difference of the $t$ and $\bar t$ production
rates at high rapidity (maybe at LHCb), where one could be sensitive to the
asymmetry without reconstructing both the $t$ and the $\bar t$ particles in the
same event.

Since neither of these measurements are straightforward, it is worthwhile to
explore possible other signatures.  One exciting possibility is that
the colored scalars discussed could manifest themselves in future
higher sensitivity dijet analyses.  The constraints are already quite
powerful for dijet resonances of valence quarks (for example
ruling out the \six). The \threebar\ appears as an $s$-channel
resonance only in  $uc \to uc$ scattering and in the
$u$-channel in $\uubar\to \ccbar$.
Depending on the choice of parameters in our model either of these
processes may be observable with increasing statistics, and would
be a spectacular discovery, especially if combined with even some
limited charm tagging.

\begin{figure}[tb]
\includegraphics[width=.5\columnwidth]{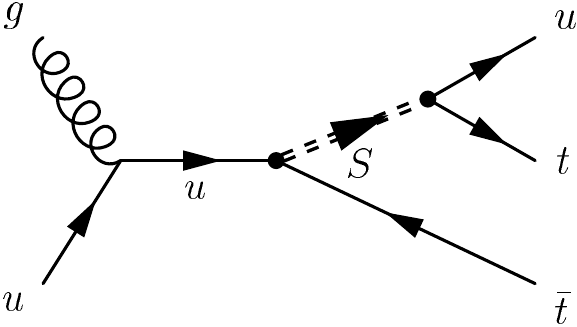} 
\caption{The flavor conserving diquark contribution to $\ttbar j$ production at
the LHC.  Two other diagrams with the gluon attached to the $S$ and to the
$\bar t$ are not shown.}
\label{fig:puppy}
\end{figure}

Another promising signature, discussed recently in Ref.~\cite{Gresham:2011dg}
(in the context of flavor violating models), is to look for $ug \to S_{ut} \bar
t \to \ttbar u$ production (see Fig.~\ref{fig:puppy}).  Since in our preferred
scenario $m_S \gg m_t$, the $u$-quark jet from the decay of $S$ would have high
$p_T^{\rm jet}$, where the standard model background (known at
NLO~\cite{Dittmaier:2008uj}) is suppressed.  The $\ttbar j$ cross section from
our model would be quite large. For example, at the 7\,TeV LHC the new physics
contribution is 2\,pb for $m_S = 600\,\GeV$ and $\lambda=1/\sqrt{2}$ (and
7.7\,pb for $m_S = 400\,\GeV$, $\lambda=1/\sqrt{2}$)~\cite{Gresham:2011dg}, and
for arbitrary coupling it is enhanced by $2\,\lambda^2$. Comparing with recent
predictions for the $\ttbar j$ distributions at the 7\,TeV LHC~\cite{alioli,
Kardos:2011qa}, we find that simply cutting on $p_T$ of the hardest jet, our
signal is somewhat smaller than the SM background.  The experimental sensitivity
can probably be optimized and substantially enhanced by measuring the $\ttbar j$
rate as a function of both $p_T^{\rm jet}$ and \mttbar\ (since for the SM
background, but not for the signal coming from $\bar t S$ decay, $p_T^{\rm jet}$
and \mttbar\ are anti-correlated), and choosing suitable cuts on both
variables. 

In addition, pair production of $S S^*$ gives rise to a \ttbar\uubar\ (or $4j$)
final state where the unflavored jets have large $p_T^{\rm jet}$. The rate is
smaller than $\bar t S$ production, however the second hard jet is advantageous
for rejecting SM backgrounds. Especially when the LHC gets to higher energy, the
sensitivity in this channel may become competitive with the $\ttbar j$ signal.

We conclude that the most promising signals of the \threebar\ at the LHC
are either the $s$-channel resonance contribution whose $uc \to uc$ origin might
become established even with very limited charm tagging, or, especially, $\ttbar
j$ production at high $p_T^{\rm jet}$ (maybe combined with an \mttbar\ cut). 
Without a much more detailed analysis than the present study, it is not possible
to determine how the sensitivities of these two searches will compare when
the actual experimental analyses are carried out.  We hope that both can be
pursued by ATLAS and CMS, as they are also interesting searches for models
beyond those motivated by the Tevatron \ttbar\ forward-backward asymmetry
discussed in this paper.

{\it Note added:} The predicted \ttbar\ cross section rises in our model compared to the SM for high invariant masses. For example, the cross section for \mttbar\ above 1 TeV at the 7 TeV LHC is larger than the QCD cross section by a factor of 2 to 3 in the preferred region of parameter space. Therefore future
measurements (combined with improved theory predictions) should discover or rule
out our model. Note that the cross section above 1 TeV due to the exchange of a
lighter particle cannot be approximated with a higher dimensional operator.
Therefore the results of \cite{Delaunay:2011gv,AguilarSaavedra:2011vw} do not
apply to our model.

\begin{acknowledgements}

We thank Nima Arkani-Hamed, Andy Cohen, Liam Fitzpatrick, Fabio Maltoni, Gilad
Perez, David Shih, Brock Tweedie, and Tomer Volansky for helpful conversations. 
This work was supported in part by the Director, Office of Science, Office of
High Energy Physics of the U.S.\ Department of Energy under contract
DE-AC02-05CH11231 (ZL) and DE-FG02-01ER-40676 (MS and GMT).

\end{acknowledgements}

\end{document}